\documentclass[9pt, a4paper, twocolumn, twoside]{pnas-new}
\templatetype{pnasresearcharticle} % Choose template 
\usepackage{graphicx, multirow}
\usepackage{xcolor}
\usepackage[]{changebar}
\usepackage{enumitem}
\usepackage{amsmath}
\setlist{noitemsep} % to leave space around whole list

\cbcolor{red}

\newcommand{\lb}{\left<}
\newcommand{\rb}{\right>}

\def\draft[#1]{\textcolor{gray}{#1}}
\def\say[#1]{\iftrue \subsubsection{#1} \fi}
\def\sep[#1]{\iffalse {\noindent {\color{blue} \small \texttt{#1} {\leaders\hbox{\rule{2pt}{0.4pt} } \hfill} }} \newline \indent  \fi}

\title{\Large{Matching in size: How market impact depends on the concentration of trading}}
\def \desniFooter {Matching in size}

\author[]{Ilija I. Zovko}
\affil[]{Aspect Capital, London \\ \vspace{0.1cm} \tiny ilija.zovko@aspectcapital.com \\ \tiny zovko@algo-quants.com}
\significancestatement{We show that filling an order with a large number of distinct counterparts incurs additional market impact, as opposed to filling the order with a small number of counterparts. For best execution, therefore, it may be beneficial to opportunistically fill orders with as few counterparts as possible in Large-in-scale (LIS) venues. 

This article introduces the concept of concentrated trading, a situation that occurs when a large fraction of buying or selling in a given time period is done by one or a few traders, for example when executing a large order. Using London Stock Exchange data, we show that concentrated trading suffers price impact in addition to impact caused by (smart) order routing. However, when matched with similarly concentrated counterparts on the other side of the market, the impact is greatly reduced. This suggests that exposing an order on LIS venues is expected to result in execution performance improvement. 
}


\begin{abstract}
\dropcap{I}n this paper we provide empirical evidence that large orders suffer greater market impact when matched with multiple trade counterparts than when matched with one or a small number of counterparts. This provides some support for using large in scale (LIS) venues to improve execution performance.  Filling an order on a venue in which a trader is able to minimise the number of counterparts to match with, all other things being equal, is expected to suffer less price impact than splitting the order and matching with many counterparts on anonymous venues.

To reach this conclusion we introduce the concept of \emph{concentration of trading} and study its effect on market impact. Trading is \emph{concentrated} if a disproportionate volume share of buying or selling in a period of time is done by a few trading firms. It is \emph{dilute} (i.e. not concentrated), if trading is spread across many firms. Buyers as well as sellers can be concentrated and will typically be a consequence of a small number of firms executing large orders.

Using a unique (but unfortunately somewhat dated) dataset from the London Stock Exchange (LSE) we are able to reconstruct the degree of concentration in buying and selling, and investigate its influence on resulting price moves. This is made possible by the record of trade matching details identifying LSE member firms participating in each trade.

Price impact, of course, is to a large degree influenced by the details of order routing, as implemented by, say an SOR, a Smart Order Router. The SOR decisions themselves can be influenced by the size of the order or the trading participation levels. Therefore, in order to isolate the effect of concentration, in our analysis we control for orderflow effects. We take into account the most common orderflow characteristics, such as the number of buyer or seller initiated trades and volumes (i.e., passive and aggressive orders), and the total number of firms buying or selling.

After adjusting for the effects of orderflow on price impact, we still find that concentrated trading results in larger impact than dilute trading.

As a practical and direct test of our findings, we show that large concentrated orders suffer noticeably less market impact in circumstances when they are matched with similarly concentrated trade counterparts on the other side of the market. This suggests that opportunistically trading large orders in size may improve price slippage.

The results presented in this paper imply that while the order routing decisions remain a key determinant of execution outcomes, it is similarly the \emph{number of counterparts} one trades with that influences performance. The fewer the number of counterparts, all other things being equal, the better the performance.
\end{abstract}

\begin{document}

\maketitle
\thispagestyle{firststyle}
\ifthenelse{\boolean{shortarticle}}{\ifthenelse{\boolean{singlecolumn}}{\abscontentformatted}{\abscontent}}{}

% \tableofcontents

\section{Introduction}
\dropcap{W}ith the introduction of MIFID II regulation, there has been an emergence of trading venues dedicated to matching in size. I.e., venues with mechanisms facilitating finding counterparts to fill an order with, in as few number of clips, or ideally in its entirety. 

While intuitively sound, the performance benefits of matching in size, in comparison to splitting an order into clips and working it on anonymous venues, are less obvious.

\sep[Introducing concentration of trading]
We address this question by introducing the concept of \emph{concentration of trading} and investigating its influence on price returns, and consequently execution performance. We start with the trivial observation that the total number of shares bought must be equal to the total number of shares sold. It is how those shares are \emph{partitioned} between the individual traders that determines trading concentration.

\emph{Concentrated buying} describes a situation where a large proportion of shares are bought by a single or a few exchange member firms. On the other hand, \emph{dilute buying} -- a term we use to describe a lack of concentration --  describes a situation where the shares bought are spread over a larger group of firms, all contributing comparable amounts to the total. Concentrated and dilute selling are defined analogously for sellers in the market. It is straightforward to see that, say, a large buy order will contribute to the concentration of the buying side of the market.

\sep[Concentration and price impact finding]
In the paper we present evidence that concentrated trading produces larger price impact than dilute trading. This means that a large order, when being matched with multiple small orders, adversely moves the market and suffers price slippage. Correspondingly, we find that when a large order is matched with a similarly large order, price impact is significantly reduced.\footnote{In the paper, we use the term `order` to denote the aggregate buying or selling by a single exchange member firm. We do not have data granular enough to recover orders originating from different trading desks, or strategies. However, one can argue that for the purposes of execution, it is the aggregate orderflow that is important anyway.}

%This in practice suggests that, if one is to minimise slippage, it may be better to match an order with one or a few counterparts, than splitting it and executing in anonymous venues. A firm executing a large order, which is contributing to a significant proportion of the total market volume, will incur larger impact cost if the order is executed against many different counterparties, than if it is executed against a single or few counterparties.

%[The below needs a revision]
%This seems in contrast to the fact that price impact \emph{per unit volume} is a concave function: per unit volume, small trades have greater impact than large trades. We will present evidence here that a potential explanation for this fact is that supply/demand is more persistent when it is concentrated than when it is dilute. This may allow market participants to detect and predict future orders resulting from the persistent supply/demand and adjust prices to profit from it.

\sep[Controlling for orderflow effects]
As mentioned, concentration is closely related to order size -- large orders tend to result in concentrated trading. Order size, on the other hand, tends to influence order routing and orderflow. It is well known that aspects of orderflow strongly affect price impact, with aggressive and passive order placement being the most exemplary one \citep{Plerou02,Berger06,Carlson06,Evans02,Payne03,Farmer05a,Kaniel08,Kumar05,Gopikrishnan00,Gabaix03a,Maslov01,Solomon01}. 
Therefore, in order to investigate the effect of concentration on price impact, we must take into account the known price effects of order routing and orderflow characteristics:
\begin{itemize}
\item the number and GBP volume of aggressive and passive trades, and
\item the number of firms buying vs. firms selling.
\end{itemize}
% We find that the effect of concentration remains relevant even after controlling for orderflow effects, contributing roughly one third in magnitude as order routing to price impact. %The results are consistent across analysed stocks and time periods (more on the robustness of the results later in the text). 

%Another mechanism relates to the fact that large trading volumes are correlated with large price moves\citep{Andersen96,Brock96,Weber06,Tauchen83,Chordia00}. 

\sep[Choice of trade aggregation time scale]
Concentration is not an instantaneous market property. Like volatility, it is an aggregate property of trading in a time window, for example a day or an hour. We have chosen to perform the analysis on daily intervals. However, the choice of aggregation remains an open question. Instead of calendar time, one can use an aggregation scheme resulting from a fixed number of trades (so called trade time), or resulting from a fixed number of shares traded (volume time). In addition to the daily aggregation, we have also ran the analysis on hourly and intervals resulting from a fixed number of trades. The main conclusions from the daily analysis are roughly borne out on the other tested aggregation schemes. While interesting in its own right, a complete study of the differences between timescales is out of scope of our current work.

\sep[LSE data that allows this analysis]
The analysis is based on a somewhat dated trades and quotes dataset from the on-book trading of the London Stock Exchange (LSE) between 2000 to 2002. What makes this dataset unique is that each trade record contains codes identifying the stock exchange member firms that participated in the trade.\footnote{Member firm codes do not identify the firms by name and are scrambled monthly and across stocks for anonymity. This makes it impossible to track firms in time or investigate the properties of their orders with good statistics.} While a member firm can make a trade both in a principal (for their own account) or agent capacity, the data shows that in more than 95\% of trades the member firms act as principal to the transaction.

We base the analysis on 32 months (674 full trading days) on 46 stocks. Excluding from the dataset days in which a particular share traded less than 500 times, yields a sample of roughly 20 million trades. 
For this set we have all the trade and quote details, including trade matching details identifying the participants in each trade. The data period is from the beginning of May 2000 until the end of December 2002. During this period, there were on average between 33 to 72 member firms trading on the on-book market of the LSE each day. The average number of firms trading increases during the period and in general there are more firms trading the more active stocks. The typical number of trades is several thousand per day for active stocks. LSE is open for trading from 8:00 to 16:30 but we have discarded data from the first and last half hour of trading to avoid possible opening or closing anomalies.

%%%%%%%%%%%%%%%%%%%%%%%%%%%%%%%%%%%%%%%%%%%
\section{Concentration of trading}
\label{section:concentration}
\begin{figure}[tb!] 
   \centering
   \includegraphics[width=0.5\textwidth]{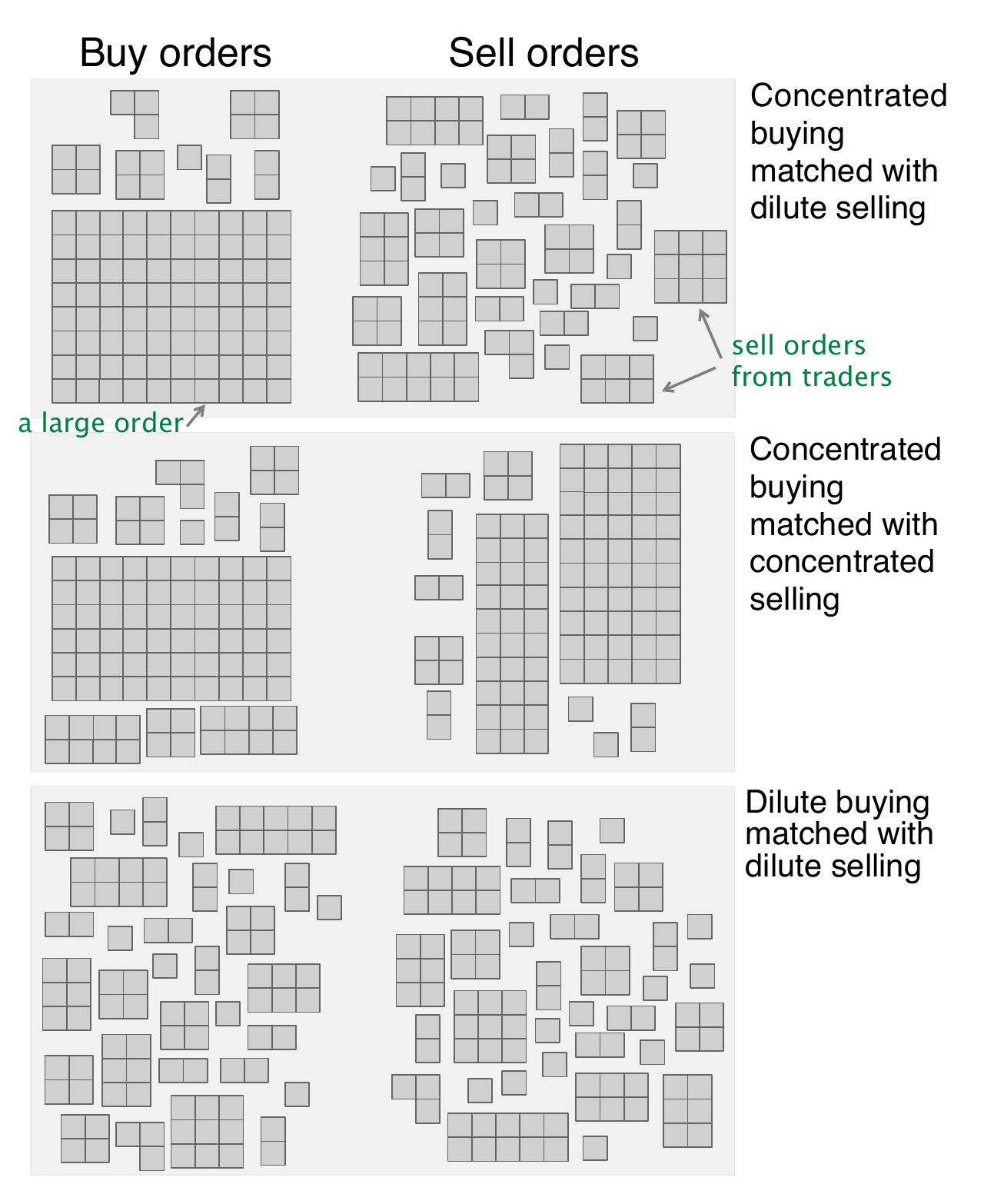}
   \caption{
Three stylised examples of trading where different levels of concentration and dilution may result in different price impact of large orders. Each small square represents a share or a ``lot'', which are then grouped by the firm that bought or sold them. In the top example, buying was concentrated as a consequence of a large order on the market. Selling, on the other hand was dilute, with a large number of orders contributing to the aggregate selling. In the middle example, both the buying and selling are concentrated. The bottom example shows trading between dilute buyers and sellers. 
\vspace{0.1cm}\\
In the paper we show that large orders causing concentration suffer adverse price impact, even after taking into account the impact of the orderflow generated by such orders. This suggests there are performance advantages to match a large order with similarly large orders, as opposed to splitting it up and filling with multiple counterparties.
%Three examples of the partition of \emph{total trade volume} into member firms' buy and sell orders (further defined in text). Small squares, representing shares or lots, are grouped by the trading firm that bought or sold them. The number of squares is necessarily equal between the buy and sell side since the number of shares bought must be equal to the number of shares sold. Their partition among the member firms however, can be different. 
%In the uppermost example, most of the buying was done by one firm while the remaining seven firms bought only a small fraction of the total trade volume. In contrast, multiple firms contributed approximately equal amounts to the selling of the shares in this example. We say that in this example the buying was concentrated, while selling was dilute; or that the demand was concentrated while supply dilute. In the mid example, both the demand and supply are concentrated in a few firms. The bottommost example shows a dilute demand and supply composition. The objective of this paper is to investigate the interaction between concentrated and dilute supply or demand on the resulting price move at the end of the trading session.
}
\label{concentrationFig}
\end{figure}

Concentrated trading is a situation where a single or a few firms contribute in a disproportionately large proportion to buying or selling in a time period. For example, a firm executing an order with a high percentage of volume (POV) algorithm will contribute to trading concentration. Alternatively, concentration can also be a consequence of a disproportionately large order being worked in the market.
Figure~\ref{concentrationFig} illustrates the principle question of the paper: do the three idealised examples have different impact on price movement, while taking into account the different orderflow characteristics expected from trading large orders.

\sep[Define volume fractions $w_i^b(t)$ and $w_i^s(t)$ and notation]
We quantify concentration using metrics of inequality between the volume fractions by which each member firm contributes to the total trade volume. Denote by $w_i^b$ the fraction of shares bought by firm $i$ in a trading session, i.e., the sum of GBP trade sizes in which firm $i$ participated as a buyer, divided by the total traded volume. Analogously, $w_i^s$ is the fraction of shares sold by firm $i$ (computed analogously). A member firm $i$, which in a given session both bought and sold, will have $w_i^s$ and $w_i^b$ non-zero.

%Let $\xi^b$ and $\xi^s$ denote respectively the sets of firms buying and selling%, and $\N^b$ and  $\N^s$ their numbers (the cardinality of sets).
%\begin{equation}
%1 = \sum_{i \in \xi^b} w_i^b = \sum_{i \in \xi^s} w_i^s.
%\end{equation}

% The total volume of shares sold in a day must of course be equal to the total volume bought. It is the partition of the buying and selling among firms -- as well as their numbers -- that varies. We can therefore partition the daily volume for buying and selling as

\sep[Quantifying concentration and its imbalance]
Volume fractions $w_i$ are positive and sum up to 1, defining a probability distribution.\footnote{We omit the superscript denoting the market side when the expression applies to both sides.} 
Hence, we can quantify concentration using any statistical measure for the inequality of a probability distribution. In order to make our results robust to the choice of the measure, we initially use two measures of inequality, the Gini index $G$ and the entropy $E$.

The sample Gini index can be computed as the half of the relative mean absolute difference between all $N$ samples $x_i$~\citep{Dixon87, Sen73}
\begin{equation}
G \equiv \frac{1}{2} \frac{\sum_{i,j} | x_i - x_j | / N^2}{\sum_i x_i/N}
\end{equation}
which for volume fractions $w_i$ can be simplified to
\begin{equation}
\label{eq:Gini}
G \equiv \frac{1}{2 N}\sum_{i,j} |w_i - w_j|.
\end{equation}
The Gini index $G_s$ for the concentration of selling is computed from the sell volume fractions, while $G_b$ for the concentration of buying from the buy fractions. 

For fully concentrated trading, in which all buys or sells are done by a single firm (having a 100\% participation ratio), the Gini index has a maximum value equal to 1. For fully dilute trading, in which all firms participate an equal amount to the total trade volume, it takes its minimum value of 0.

The entropy based measure for concentration is defined as
\begin{equation}
\label{entropy}
E \equiv 1 - \frac{\sum_i w_i \cdot \log(1/w_i)}{\log N},
\end{equation}
where we have changed the sign and introduced the normalisation factor of $\log N$ to the usual entropy expression. With these changes, we get the intuitive interpretation that large concentration is associated with large values of the metric, consistent with the Gini metric: for fully concentrated trading the entropy measure is equal to 1, while for fully dilute it is 0.

Finally, we define the \emph{imbalance} in trading concentration between the buying and selling as
\begin{eqnarray}
\label{eqImb}
\delta G & \equiv & G_b - G_s \;\; \text{and} \nonumber \\ 
\delta E & \equiv & E_b - E_s.
\end{eqnarray}
If the buying is more concentrated than selling, the imbalance will be positive.

\sep[How often concentrated trading happens]
One might immediately wonder how often do we see trading situations in which either buying or selling is significantly concentrated. Looking at all stocks in our LSE dataset, on almost half of the days there was at least one firm that traded more than a quarter of all the buying or selling on the day. So a fairly frequent occurrence.

%%%%%%%%%%%%%%%%%%%%%%%%%%%%%%%%%%%%%%%%%%%%%%%%%%%%%%%%%%
\section{Concentration and price returns}
\label{section:concentrationAndpriceReturns}
\sep[Price return model]
%In figure~\ref{concentrationFig}, we have illustrated three idealised situations in which we will aim to understand the potential impact of concentration on price moves. Consequently we will show that large orders contributing to the concentration suffer price impact in addition to the known impact of orderflow.
We have decided to estimate the effects of concentration and orderflow on price returns using a linear model. After trying out a few nonlinear forms, we did not find clear evidence of nonlinearities. The extremes of the dataset do deviate from the linear assumption, but the noise levels are large and do not warrant additional complexity. We have also found that the Entropy and Gini metrics bear same conclusions, so will restrict the discussion to only using the Entropy metric. 

Therefore we will be explaining the daily price returns $\delta P_t$ by expressions of the form
\begin{eqnarray}
\delta P_t &=& \text{(concentration imbalance)}_t \nonumber \\
			&+& \text{(order routing imbalances)}_t \\
			&+& \text{(error term)}_t. \nonumber
\end{eqnarray}
Price returns are calculated as the percent difference of the volume weighted average price (VWAP) of the last 10\% and first 10\% of trades on the day\footnote{Recall that we discarded 30 minutes of data after the opening auction, and 30 minutes before the close.}. We normalised the returns by the overall market return, subtracting from each daily stock return the FTSE100 return for the day. To make sure that overall stock moves do not impact conclusions, mean return per stock over the entire period is subtracted from the daily stock return. The variance of price returns across stocks is left unchanged.

\sep[Orderflow imbalances and stock normalisation]
The order routing effects are captured by the imbalances between buying and selling in
\begin{itemize}
\item numbers of aggressive orders $M_t$,
\item GBP volume of aggressive orders $V_t$, and
\item numbers of firms $N_t$.
\end{itemize}
The imbalances are computed using the usual convention that buying gets a positive sign, and using the common normalising form
\begin{equation}
\delta x = \frac{x_b - x_s}{x_b + x_s}
\label{imbForm}
\end{equation}
where $b$ and $s$ designate buying and selling. The full orderbook data we use allows us to calculate these order routing variables precisely.

In summary, for each stock and trading day, we compute the market normalised price return $\delta P_t$ and the concentration imbalance between buyers and sellers $\delta E_t$ using Eq.\ref{eqImb}. Using Eq.\ref{imbForm}, we compute the imbalance between the number and GBP notional of aggressive buy orders and aggressive sell orders, $\delta M_t$ and $\delta V_t$ respectively. And similarily the imbalance in the number of buyers and sellers $\delta N_t$. To combine data from different stocks, each imbalance variable is further normalised by the standard deviation for the corresponding stock.

%:figure orderflow correlations
\begin{figure}[t!]
\begin{center}
	\includegraphics[scale=0.275]{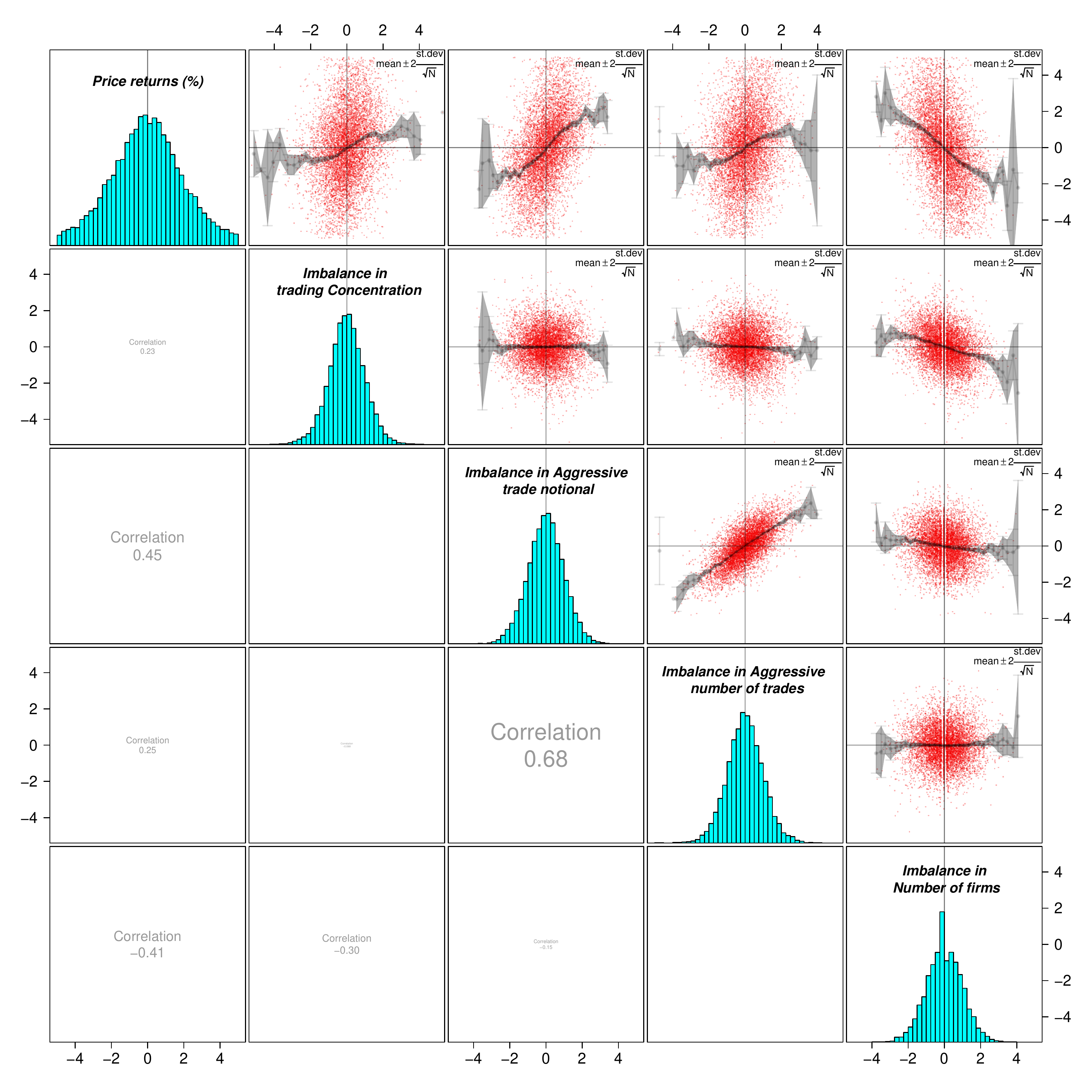}
	\caption{Scatterplots, histograms and correlations showing the dependency between price returns and imbalances in concentration and orderflow. 
	Matrix diagonal shows the histograms, roughly normally distributed with mean 0 and standard deviation 1. The exception is the distribution of price returns, which in fact are not normalised - price returns are left as percentages. The triangle above the diagonal shows simple scatterplots of daily values (in red), together with a simple binned conditional mean estimate. Values in the matrix below the diagonal are the values of the correlation coefficients between the corresponding variables with the font size proportional to the absolute value of the coefficient. \vspace{0.1cm} \\
	The charts show a fairly correlated set of variables, which are all in one way or another related to trade volume. It is for this reason that we will have to take special care to properly account for the effects of orderflow in revealing the role of concentration in price returns.
	}
\label{matrix}
\end{center}
\end{figure}

\sep[Collinearity and regression robustness checks]
Figure~\ref{matrix} shows a matrix of dependencies between the price returns and the imbalance metrics. The upper matrix triangle shows scatter plots (in red) and a simple conditional mean estimate as the solid gray line. The level of correlation (estimated coefficients shown in the lower matrix triangle) is high since none of the orderflow variables can vary fully independently of the others. If a new buyer starts trading, changing the imbalance in the number of firms, this may in turn change the ratio of aggressive sell and buy orders in the market. Depending on the order size, it may also change the concentration of buying. 

%Notably, the imbalance in the number and the notional of aggressive trades are highly correlated. At almost 70\% correlation, they capture the same effect as we will see later. 

Trading concentration, on the other hand, is fairly independent and only mildly anti-correlated with the imbalance in the number of firms, e.g., the smaller the number of sellers, the more concentrated selling is. The fact that concentration and orderflow are rather independent is indicative of fairly sophisticated SORs at work. Even when there is a large order being worked (resulting in concentrated trading), the order routing does not to create biases in orderflow.

%%%%%%%%%%%%%%%%%%%%%%%%%%%%%%%%%%%%%%%%%%%%%%%%%%%%%%
%:Regression table
\begin{table}
\begin{center}
\scriptsize
\begin{tabular}{|l| r c c c |}
\hline
{\tiny Number of samples = 15491}				& Coef.		 &Error	&$R_S^2$	&$R_P^2$	\\
\hline
$\delta E$ (trading concentration imbalance)	&\bf  24.9   &1.3  	&0.05		&0.02		\\
$\delta N$ (imb. in num. of firms trading)		&\bf -61.2	 &1.4  	&0.17		&0.16		\\
$\delta V$ (imb. in aggressive notional)		&\bf  81.8	 &1.8  	&0.20		&0.12		\\
$\delta M$ (imb. in aggressive trade cnt.)		&    -3.0	 &1.8  	&0.06		&0.00		\\
\hline	
Overall 										&\multicolumn{4}{r|}{$R^2=0.33$}	\\
\hline
\end{tabular}
\end{center}
\caption{Model fit showing the magnitude and significance of concentration and order routing imbalances on price returns (in basis points). The impact of concentration on the price is adverse, in that positive imbalances corresponding to concentrated buying are associated with positive price returns. The impact of order routing is as expected: an imbalance towards aggressive orders adversely affects the price, as well as the fewer firms trading on one side of the market, the more adverse price impact they can expect, reflecting their order size. Only the imbalance in the number of aggressive trades is not significant, with a similar effect better captured by aggressive notional imbalance. %The constant in the regression is not significant because the returns were demeaned.
\vspace{0.1cm} \\
The effect of concentration is comparable in magnitude ($\text{Coef} \approx$ 25 bps) to the impact of routing ($\text{Coefs} \approx$ 80 and -60 bps), but the explanatory power of concentration imbalance is less than that of order routing ($R^2_p \approx$ 0.02 vs. 0.16 and 0.12). In other words, impact predictions based on trading concentration are equally important, but are less reliable and more noisy. $R_P^2$ is the partial $R^2$ of the selected variable. It expresses how much the variable explains price returns in addition to the other three variables. $R_S^2$ is the value of $R^2$ in a regression where only the selected variable is present in the regression. It expresses how much the variable on its own explains price returns.}
\label{tabReg}
\end{table}
%%%%%%%%%%%%%%%%%%%%%%%%%%%%%%%%%%%%%%%%%%%%%%%%%%%%%%%%

\sep[Model and fit, $R^2$]
Ultimately the model to estimate is
\begin{equation}
	\label{model}
	\delta P_t = \alpha \cdot \delta E_t \; + \; \beta \cdot \delta M_t + \gamma \cdot \delta V_t + \tau \cdot \delta N_t \; + \epsilon_t , 
\end{equation}
where $\epsilon_t$ are $\mathcal{N}(\mu,\sigma)$ Normal residuals. We estimate the parameters via a standard OLS on a merged dataset for all stocks, containing 15491 samples corresponding to day sessions. In doing this, we have limited ourselves to sessions containing more than 500 trades and have removed sessions in which the stock price changed by more than 5 percent (to limit the effect of exogenous or news events). The number of samples we removed is around 5\% and in reality does not materially change the conclusions. In terms of other robustness checks, we performed the analysis in distinct yearly periods as well as separately for each stock, broadly corroborating the conclusions.

Model estimates and corresponding errors are collected in table~\ref{tabReg}. The model explains about 30\% of overall price return variation (model $R^2=0.33$). However, if we focus on large returns and estimate the model on days where there was a price change larger than 1\%, the explanatory power increases to 40\% ($R^2=0.40$). If we further restrict the sample to include only returns between 4\% and 5\%, the explanatory power increases further to 60\%. In general, the model explains larger price returns better than smaller.

In the table we also calculate $R_S^2$ and $R_P^2$. These goodness-of-fit measures describe the contribution of each of the imbalances to the overall model fit. $R_S^2$ is equal to the r-squared of a model with \emph{only the selected} variable, and no others, included\footnote{As such, it is equal to the square of the correlation between the variable and price returns, shown in figure~\ref{matrix}.}. $R_P^2$ is the partial r-squared defined as $R_P^2 = (R^2 - \hat{R}^2)(1-\hat{R}^2)$ where $R^2$ is the R-squared of the full model with \emph{all} variables included, and $\hat{R}^2$ is the R-squared of a restricted model with \emph{all except} the selected variable. $R_P^2$ is the partial contribution to the full r-squared due to the selected variable.

\sep[Regression discussion]
Not surprisingly, order routing explains a large proportion of variation in price returns (large $R^2_p$). However, that does not mean that order routing is the single \emph{``important''} contributor to market impact. Loosely speaking, \emph{``importance''} of a factor is a combination of how \emph{large} an effect the factor predicts, and how \emph{reliable} a prediction it produces. I.e., an unreliable prediction of a large magnitude may be more ``important'' than a reliable prediction of a small inconsequential effect. $R^{2}$ describes how reliable a factor is in predicting the price impact, while the estimated coefficient determines the magnitude of the effect. 

In the estimated model, price returns were expressed in basis points. The concentration and orderflow imbalances are without units, however they are normalised by the stock standard deviation. Hence the units of regression coefficients can be read as basis points per unit of typical imbalance. Therefore, while the imbalances in order routing and the number of firms trading explain respectively around $R_p^2(\text{market order volume}) = 0.12$ and and $R_p^2(\text{number of firms}) = 0.16$ to the overall $R^2$. The magnitude of their effects for one standard deviation are estimated to be 80 bps and -60 bps respectively. The contribution of concentration imbalance, is more noisy with $R_p^2(\text{concentration}) = 0.02$, but has a comparable 25 bps effect on the magnitude of the price impact.

\sep[Model illustration]
As an illustration we can take a sample from the dataset, for example LLOY on $t=\text{2000-05-09}$. The market adjusted return on that day was $\delta P_t = -1.8\%$.
Explanatory variables on the day were:
\begin{equation}
\delta E_t = -2.9 \hspace{0.3cm} \delta V_t = -0.47 \hspace{0.3cm} \delta N_t = 1.36 \hspace{0.1cm}. \nonumber
\end{equation}
We omit aggressive trade count because its contribution is not significant and likely captured by the aggressive notional. On the day, the concentration of sellers was about 3 times as large as typically, and there was a slightly higher number of buyers than sellers. The number of aggressive sell orders (market orders) was not notably larger than the number of aggressive buy orders. Upon multiplying these values by the coefficients and adding up, the predicted return for the day is
\begin{align}
\delta P_t 		&= 25 \cdot \delta E_t 	+80 \cdot \delta V_t  -60 \cdot \delta N_t  \nonumber \\
-200 \text{bps} &= -99 					-23 				  -77.
\end{align}
We see that the contribution of concentration to the predicted return is of comparable magnitude as orderflow, albeit with lower reliability. Across the sample of all stocks, the contribution of concentration is between 20\% and 30\% of the total return.
%Imbalance in buying vs. selling concentration was $\delta E_t = -2.9$; in aggressive trade notional $\delta V_t = -0.47$; in the number of buyers and sellers $\delta N_t = 1.36$; we omit aggressive trade count because it's contribution is not significant and likely captured by the aggressive notional. 

\sep[Model specification]
Checking the statistical specification of the model, the residuals are i.i.d.\ and very close to normal; all the explanatory variables are exogenous to the model. Therefore, the model coefficients are estimated without bias and are distributed according to a normal distribution. The reported regression coefficient errors and the p-values are calculated in the standard way assuming normality. However, we have also confirmed the estimation errors using a bootstrap test.\footnote{By shuffling the price returns and keeping all other variables intact, we obtain a realisation of the null hypothesis where all the explanatory variables are correlated with themselves but are uncorrelated with the returns. Repeating the shuffling 1000 times and estimating the model on the bootstrapped data, we get a distribution of the coefficients under the null. The standard deviation of the estimates and the p-values obtained in this way coincide with the theoretical values shown in the table.} 
Recursive and sliding window estimates are all in line with the overall model fit. In addition, we split the model by year and by stock name. The estimated values are rather constant over the years and do not greatly vary between stocks. We can note however, that higher activity stocks tend to result in better model fits.

\sep[Orthogonality of concentration to order routing]
While the model and the estimates are reliable, a complication in the interpretation of results is that all of the explanatory variables are, in one way or another, driven and influenced by traded volume. To determine that trading concentration influences price moves in its own right, it is important to determine that the effect of concentration is independent or orthogonal to order routing.

One way of establishing that this is the case, is to split the single model Eq.~\ref{model} into, so called, partial regressions, and estimate them via a two stage process. In the first stage we remove the linear effects of order routing from both the price returns \emph{and} trade concentration by fitting models 
\begin{eqnarray}
	\delta P_t &= \alpha_1 \cdot \delta V_t + \beta_1 \cdot \delta N_t + \gamma_1 \cdot \delta M_t + \epsilon_{1,t} , \nonumber \\
	\delta E_t &= \alpha_2 \cdot \delta V_t + \beta_2 \cdot \delta N_t + \gamma_2 \cdot \delta M_t + \epsilon_{2,t}.
\label{partialReg}
\end{eqnarray}
The residuals of these fits, $\delta P'_t \equiv \hat{\epsilon}_{1,t}$ and $\delta E'_t \equiv \hat{\epsilon}_{2,t}$, we can term \emph{routing corrected price returns} and \emph{routing corrected concentration}. The second step entails regressing the ``corrected'' variables on each other
\begin{equation}
\delta P'_t = \eta \cdot \delta E'_t + \epsilon_{3,t}.
\label{residReg}
\end{equation}
This procedure removes linear effects of order routing from price returns and trading concentration imbalance. Whatever significance remains in the second step is the effect of trading concentration on price returns  -- orthogonal to order routing. 

\begin{figure}[t!]
\begin{center} 
	\includegraphics[scale=0.4]{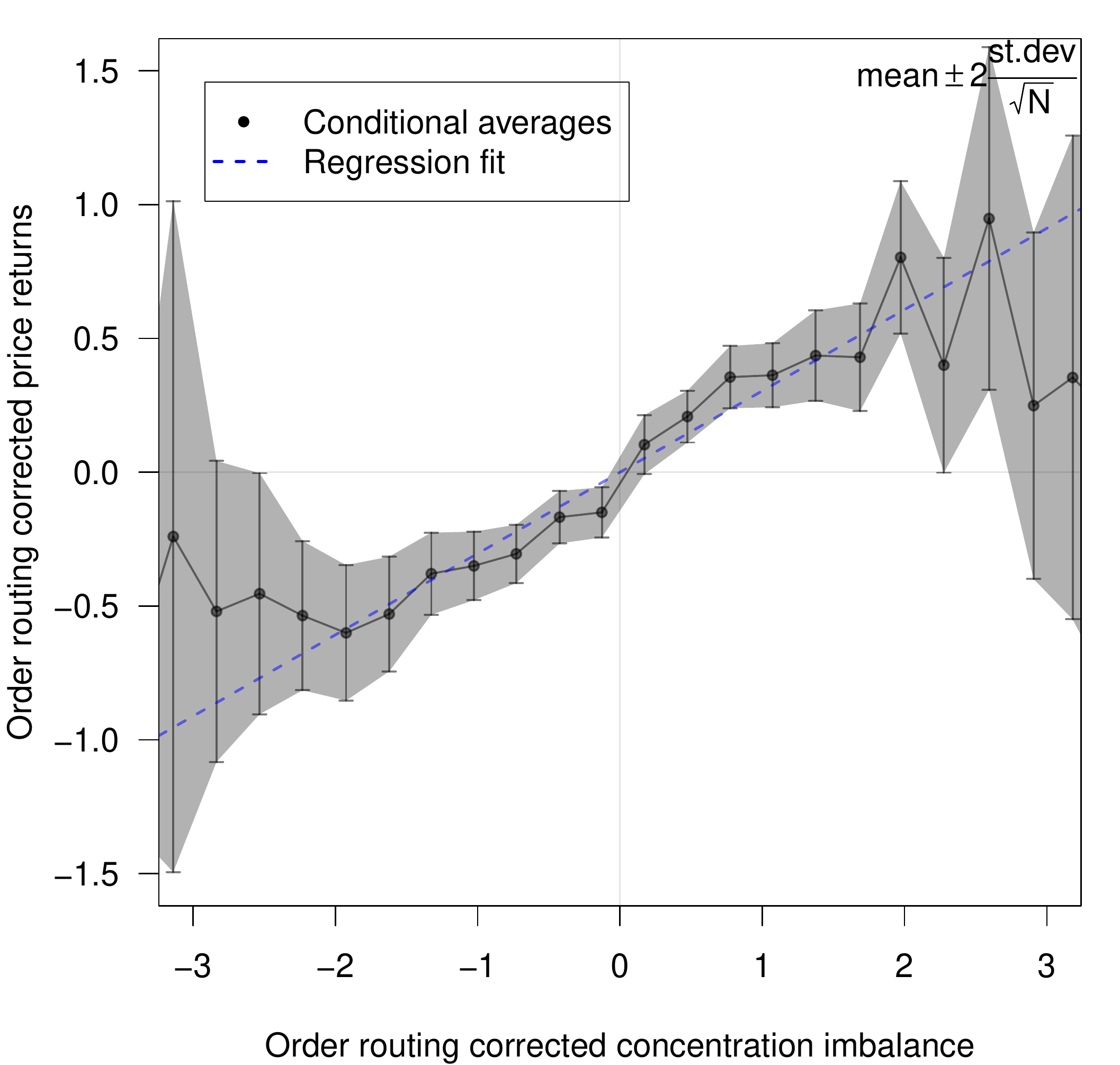}	
	\caption{Figure showing the significant relation of concentration and price impact, after taking order routing effects into account. Days where concentration was high on one side of the market tend to have larger adverse price moves in stock price. \vspace{0.1 cm} \\
	On the x-axis we plot the residual information in the concentration imbalance, once orderflow effects have been regressed out. This is comptuted as the residuals of the concentration regression in the first of the two stage partial regression process described in the text ($\epsilon_{2,t}$ in Eq.\ref{partialReg}). Likewise, y-axis shows the residual price return $\epsilon_{1,t}$, after removing the effects of orderflow imbalances. Points in the chart are binned variables averages, the line is the regression line from Eq.\ref{residReg}. 
}
\label{priceDelta}
\end{center}
\end{figure}

A fit of Eq.~\ref{residReg} yields $\hat{\eta}= 0.25 \pm 0.01$ with p-value zero and $R^2=0.02$, very much in line with the results in table~\ref{tabReg}. Fig.~\ref{priceDelta} shows the fit graphically, together with simple conditional averages and the corresponding standard normal errors of the residuals.

What this shows is that the influence of trading concentration on market returns is indeed orthogonal to order routing, and typically about one quarter in magnitude ($\hat{\eta}= 0.25$). %The problem of optimal routing and scheduling is a multivariate problem, in which all the variables need to be considered together. However, trading concentration should be included is such an optimisation.

%:Figure partial regs

\section{Executing large in scale}
The preceding section showed that trading concentration \emph{on average} adversely impacts price moves. Practically speaking however, if one was to trade a large order, what might be the ways of mitigating the additional impact due to concentration? LIS venues aim to reduce impact by matching offsetting interest only when it can be matched in size between a few counterparts. In other words, matching only when both the buy side and the sell side are concentrated. By appropriately partitioning the LSE data we can test this approach and determine if matching concentration with concentration reduces impact.

To do this, we partition the samples into categories, each indexing days from one of the regimes corresponding to illustrations in Fig~\ref{concentrationFig}:
\begin{itemize}
\item Concentrated buying matched with concentrated selling;
\item Concentrated buying matched with diluted selling;
\item Dilute buying matched with concentrated selling;
\item Dilute buying matched with dilute selling.
\end{itemize}
Once we remove the effects of order routing, we are left with the average price move conditional to the four concentration regimes. A way to compute these conditional price moves is to create four, so called, ``dummy'' variables indexing the four market regimes. The model is very much like Eq.~\ref{model}, with the difference that the  continuous concentration imbalance $\delta E_t$ is replaced with the four categorical variables. In defining the categories, we consider a market side to be concentrated if the concentration measure is larger than the 70th quantile of the metric, and dilute if it is smaller than the 30th quantile.\footnote{To estimate the quantiles, we merge the concentration metrics for the sell and buy sides.}

%%%%%%%%%%%%%%%%%%%%%%%%%%%%%%%%%%%%%%%%%%%%%%%%%%%%%%%%%%%%%%%
%:Table dummy regressions - info content
\begin{table}[tb]

\scriptsize
\begin{center}
\begin{tabular}{|l| r c c |}
\hline
\bf									& Coef.			&Error	&p-val  		\\
\hline
Signed volume, $\delta V$			&\bf	 82.1	&1.8	&0.00	\\
No.\,firms, $\delta N$				&\bf	-65.4	&1.4	&0.00	\\
No.\,signed trades, $\delta M$		&		 -4.4	&1.8	&0.01	\\
Concentrated sell, dilute buy		&\bf	-44.0	&4.8    &0.00	\\
Concentrated buy, dilute sell		&\bf	 33.0	&4.8    &0.00	\\
Dilute sell, dilute buy				&		 -3.8	&4.3	&0.38	\\
Concentrated sell, concentrated buy &		  4.1	&4.3	&0.34	\\
\hline	
Overall &\multicolumn{3}{r|}{$R^2= 0.33$}							\\
\hline	
\end{tabular}
\end{center}

\caption{Regression results showing the different effects of concentration on the resulting price move split by different levels of concentration on the two market sides. As before, the concentrated side of the market suffers adverse price moves, but only when matched with a dilute opposite side.  When a concentrated order is matched with a concentrated counterpart, the market impact caused by concentration vanishes.
\vspace{0.1cm} \\
The orderflow effects ($\delta V, \delta N, \delta M$) and coefficients are largely unchanged from before. Likewise, concentrated selling when trading with diluted buying results on average in a price drop ($\text{coef}=-44 \pm 5 \text{bp}$); concentrated buying when matched with dilute selling results in price appreciation ($\text{coef}=33 \pm 5 \text{bp}$). However, when concentrated trading is matched with similarity concentrated counterparts, the impact of concentration vanishes. This observation leads us to expect performance improvements for algos utilising logic to restrict the number of counterparts to trade with.
}

\label{dummy}
\end{table}
%%%%%%%%%%%%%%%%%%%%%%%%%%%%%%%%%%%%%%%%%%%%%%%%%%%%%%%%%%%%%%%%%%%%%%%%

\sep[Dummy regression fit]
Table~\ref{dummy} displays the model fit containing the four dummy variables. The effect of order routing remains significant and very similar to previously observed. What is interesting here is that of the four conditional means only the ones indexing concentrated with dilute trading are significantly different from zero. Additional price impact due to concentration vanishes in situations where concentrated trading is matched with concentrated counterparts.

% It is only situations with one side of the market being concentrated that lead to an increase in market impact. If both sides of the market are similarly concentrated, i.e., in a situation where large sellers trade with large buyers, there is no additional price impact.

\sep[Searching large counterparts]
Given the above, it seems beneficial (if constraints of the execution allow it) to fill large orders with similarly large opposite interest, minimising the number of trade counterparts. In an anonymous, centrally cleared market it may not be possible to know directly the number of distinct counterparts an order is filled with. However, while far from straightforward, from fill periodicity, size or other patterns, there may be ways to indirectly infer the information. Algos with such logic could, for example, speed up the execution to profit from times when trading with few concentrated counterparties, and slow down when trading with many counterparts.

The other possibility is to advertise interest in large-in-scale (LIS) venues. Arguably, the upstairs voice market of the LSE is a venue designed with such a purpose. The brokers would dial their contacts in search for a few counterparts to fill the order in size. However, trading in the upstairs market brings with it information leakage complications. LIS venues may achieve a similar purpose, but with information leakage tightly controlled and measurable.

\section{Signalling and time persistence of concentration}
\label{section:signalling}
This analysis is based on trading in an anonymous and centrally cleared market in which information leakage is minimised. In spite of this, we speculate that it likely is information leakage that is responsible for the observed impact of concentrated trading.

Market impact is known to be a convex function of order size~\citep{Potters03, Lillo03a, Farmer05a}, a fact we were able to confirm on the LSE dataset we use as well. The implication of this is that impact per share (or per notional) is larger for small orders than it is for large ones. From a pure mechanical impact view, therefore it follows that when a large order is matched with many small orders, the sum of impacts of small orders should be larger than the impact of the large order. In the language of concentration, the many small orders on the dilute side of the market should -- in aggregate -- cause more price impact than the large order on the concentrated side of the market. From earlier results, we know the contrary to be the case.  Clearly, a simple mechanical explanation does not capture the full story.

\begin{figure}[t] 
\begin{center} 
	\includegraphics[width=0.5\textwidth]{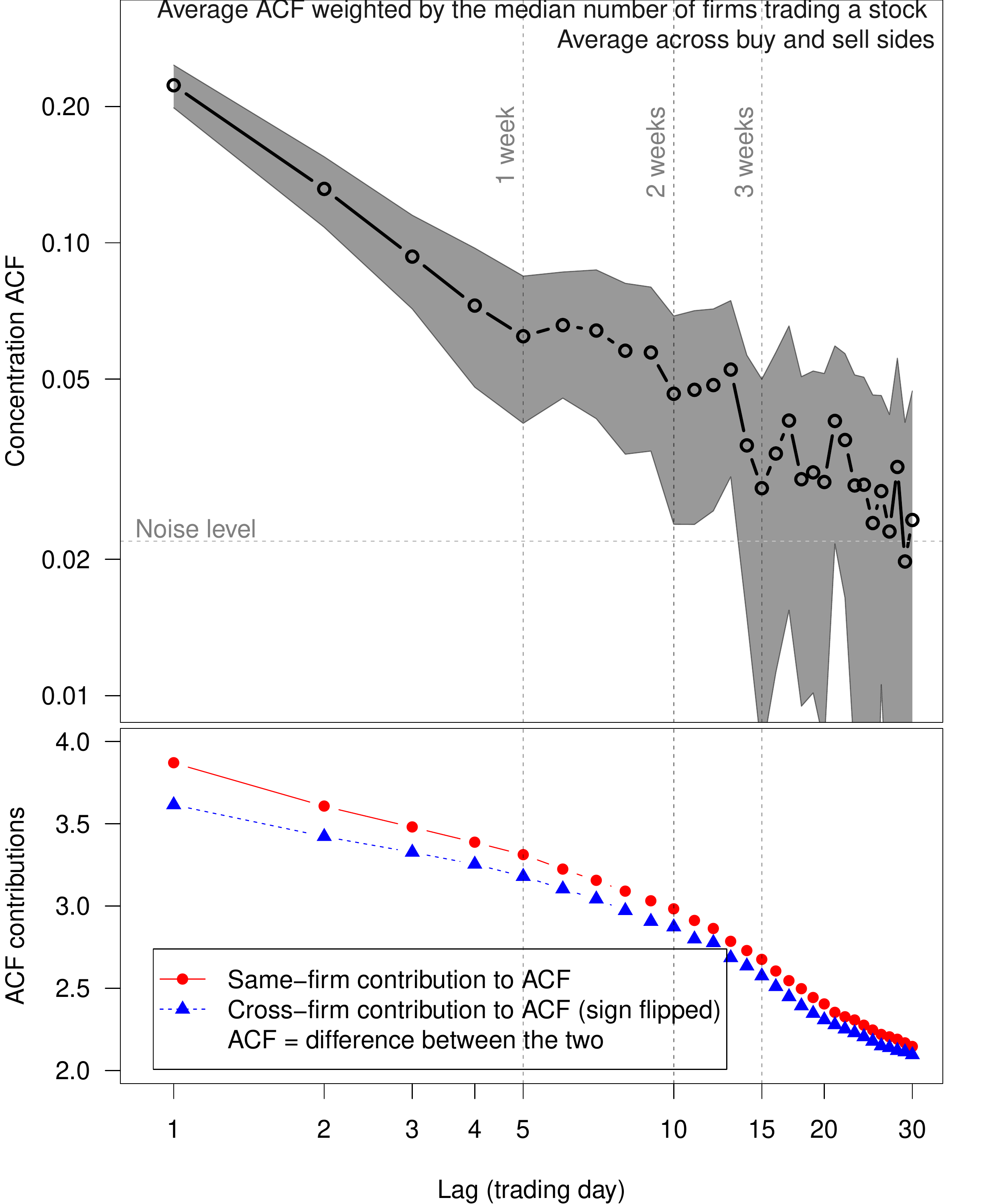}	
	\caption{Autocorrelation function (ACF) of trading concentration showing strong persistence over days (upper chart). We speculate it is this persistence in concentration that allows the dilute side of the market to adjust their behaviour and profit from the knowledge of a large order contributing to concentrated trading. The decay seems to roughly be a power curve (note the log-log scale), and show strong persistence up to a week. The result is obtained by averaging across stocks and both market sides. \vspace{0.1 cm} \\
The lower chart shows that the persistence in concentration is caused by the same-firms executing large orders across multiple days. We break the aggregate ACF into same- and cross-firm correlations. The effects of the two are opposite, with the same-firm contributing positively to correlations. (We flipped the sign of cross-firm correlations so we can chart it better). The difference between the two curves is the resulting aggregate ACF.
}
\label{acfFig}
\end{center}
\end{figure}

We speculate it is likely that signalling and information leakage by the large order allows the ``dilute side`` firms to profit by adjusting prices as the intent and size of the large order is revealed to the market. For them to be able to do so, a certain level of persistence in concentration is required. Figure~\ref{acfFig} shows that, indeed, there is significant autocorrelation in the levels of concentration from one day to the next. Correlations decay roughly as a power function, are strongly significant up to a week, and weakly for multiple weeks. Even the ACF of the concentration imbalance is significant up to a few days.

\sep[Origins of correlations]
%:Figure concentration ACF

Correlations in trading concentration can be generated in two ways: (i) one or a few firms trading large orders across days, or (ii) different firms entering the market with large orders sequentially.

The former, ``same-firm'' correlations, are typically a consequence of splitting large orders and trading them sequentially~\cite{Lillo04a, Bouchaud04}. The latter, ``cross-firm'' correlations can be caused by news events and its different propagation among firms, or by the so called ``herding hypothesis``~\cite{Lakonishok92} whereupon observing large orders, traders respond by placing own large orders. A ``hot-potato`` is another variant of the herding hypothesis in which traders sequentially trade a large position between themselves~\cite{Lyons97}.
 
By splitting the empirical ACF into two components, one contributed by same-firm correlations, and another with cross-firm correlations, we can determine which of the two mechanisms contributes to the persistence in concentration.

\sep[ACF components]
The autocorrelation function $\gamma(\tau)$ is defined as
\begin{equation}
\gamma(\tau) = \frac{ \lb E(t) \cdot E(t+\tau)\rb}{\lb E(t) \cdot E(t) \rb}
\label{acf}
\end{equation}
where the bracket notation $\lb \cdot \rb$ denotes averaging over days\footnote{This expression holds for zero mean variables. Average entropy is not zero, hence we will need to remove the mean prior to computing the decomposition.}. We separate the contributions to the ACF by introducing summands $z_i(t)$ so that the concentration on a given day can we written as
\begin{equation}
\label{entEq}
E(t) = \sum_{i \in \xi_t} \frac{- w_i(t) \log(w_i(t))}{\log N(t)} \equiv \sum_{i \in \xi_t} z_i(t).
\end{equation}
Substituting the definition of concentration into the expression for the ACF we obtain
\begin{equation}
\gamma(\tau) = \frac{\lb \sum_i z_i(t) \cdot \sum_j z_j(t+\tau) \rb}{\lb \sum_i z_i(t) \cdot \sum_j z_j(t) \rb}
\end{equation}
which can be rearranged as
\begin{equation}
\gamma(\tau) = \frac{\lb \! \sum_{i=j} z_i(t) \! \cdot \! z_j(t\!+\!\tau) \! \rb + \lb \! \sum_{i \neq j}z_i(t) \! \cdot \! z_j(t\!+\!\tau) \! \rb} {\lb \sum_{i,j} z_i(t) \cdot z_j(t) \rb}.
\end{equation}
The first sum takes into account the contributions to the ACF from same-firm, while the second term takes into account cross-firm contributions. It turns out that the second sum is negative, so we will for convenience change the sign of the second sum and write the ACF as 
\begin{equation}
\gamma(\tau) =  \gamma_{\text{same}}(\tau) - \gamma_{\text{cross}}(\tau).
\label{gamma}
\end{equation}

\sep[Demeaning the concentration measure]
Prior to decomposing the ACF however, we need to ensure $E(t)$ has zero mean which we do by subtracting the mean  
\begin{equation}
E'(t) \equiv E(t) - \overline{E} = E(t) - \frac{1}{T} \sum_d E(d).
\end{equation}
As before, denoting by $\xi_d$ the set of firms or orders present in the market on day $d$, we can write the above expression as
\begin{eqnarray}
E'(t) & = & \sum_{i \in \xi_t} z_i(t) - \frac{1}{T} \sum_d \sum_{j \in \xi_{d}} z_j(d) \nonumber \\
	  & = & \sum_{i \in \xi_t} \left( z_i(t) - \frac{1}{N_t} \overline{E} \right) \equiv \sum_{i \in \xi_t} z'_i(t)
\end{eqnarray}
where $N_t$ again stands for the cardinality of the set $\xi_t$. %and $\overline{E} \equiv \frac{1}{T} \sum_d \sum_i z_i(d)$. 
We can now compute the components of the ACF using $z'_i(t)$.

\sep[Discussion]
The analysis reveals that same-firm and cross-firm correlations are of a different sign and work in opposite directions (lower panel of Fig~\ref{acfFig}). Same-firm correlations increase overall ACF and are very closely matched in magnitude by the reduction due to the anti-correlation of cross-firms. The small difference in magnitude between the opposing forces results in the overall concentration ACF.

While this analysis does not rule out that the ACF decomposition is purely a mechanical effect due to how we compute entropy, it does show that the persistence in concentration is due to large order splitting. As such it fits well in the common narrative that large orders leak information during execution.

\section{Conclusions}
\label{section:conclusions}
This paper introduces the concept of trading concentration and investigates its effect on price impact. When a firm executes a large order, in contributes to the concentration of that side of the market. In such a situation, in addition to price impact due to the way how the order is traded (SOR order routing), concentration will have an adverse effect on price impact, in magnitude roughly 30\% as the effects of orderflow. A way to reduce this additional impact is to selectively fill the order with comparably large opposing interest.

This can either be facilitated by expressing interest in LIS venues, or by inferring patterns of fills when trading in an anonymous market, which may signal the presence of large opposite interest. In such situations, it is advantageous to speed up trading to minimise the number of trade counterparts.

We speculate that large orders, spread over hours or days, over time leak their intent even when trading in anonymous markets. This allows smaller traders to adjust prices and profit from the large order fills.

This paper does not unequivocally find support for trading in large in scale venues. This would be impossible only with data from anonymous trading. What we do show is that even in a standard continuous double auction market, taking into account orderflow effects, price impact of an order is increased when the order is filled with a large number of counterparts.

\color{black}
\bibliographystyle{unsrtnat}
\bibliography{iiz10}

\acknow{The author thanks J.Doyne Farmer for providing the LSE data used in the analysis and early discussions. Thanks to Roel Oomen for reading a draft and providing useful comments.}

\showacknow{} % Display the acknowledgments section

\end{document}